\documentclass[preprint,12pt,3p]{elsarticle}

\usepackage{amssymb}
\usepackage{CJK}
\usepackage{stfloats}
\usepackage{amsmath,amssymb,amsfonts}
\usepackage{algorithmic}
\usepackage[colorlinks,linkcolor=blue]{hyperref}
\usepackage{graphicx}
\usepackage{textcomp}
\usepackage{xcolor}
\usepackage{algorithm} 
\usepackage{multirow} 

\journal{Computers $ \& $ Security}

\begin{document}
	
	\begin{frontmatter}
		
		\title{HSTF-Model: an HTTP-based Trojan Detection Model via the Hierarchical Spatio-Temporal Features of Traffics}
		
		\address[label1]{Institute of Information Engineering, Chinese Academy of Sciences, Beijing, China}
		\address[label2]{Key Laboratory of Network Assessment Technology, University of Chinese Academy of Sciences, Beijing, China}
		\address[label3]{School of Cyber Security, University of Chinese Academy of Sciences, Beijing, China}
		\address[label4]{National Computer Network Emergency Response Technical Team/Coordination Center of China, Beijing, China}

		\author[label1,label3]{Jiang Xie}
		\ead{xiejiang@iie.ac.cn}

		\author[label1,label2]{Shuhao Li\corref{cor1}}
		\cortext[cor1]{The corresponding author of this paper is Shuhao Li. This work is supported by the National Key Research and Development Program of China
			(Grant No.2016YFB0801502), and the National Natural Science Foundation of China
			(Grant No.U1736218). }
		\ead{lishuhao@iie.ac.cn}
		
		\author[label1,label2,label4]{Xiaochun Yun}
		\ead{yunxiaochun@cert.org.cn}
		
		\author[label1,label2,label3]{Yongzheng Zhang}
		\ead{zhangyongzheng@iie.ac.cn}
		
		\author[label1,label2]{Peng Chang}
		\ead{changpeng@iie.ac.cn}
		
		\begin{abstract}
			HTTP-based Trojan is extremely threatening, and it is difficult to be effectively detected because of its concealment and confusion. Previous detection methods usually are with poor generalization ability due to outdated datasets and reliance on manual feature extraction, which makes these methods always perform well under their private dataset, but poorly or even fail to work in real network environment. In this paper, we propose an HTTP-based Trojan detection model via the Hierarchical Spatio-Temporal Features of traffics (HSTF-Model) based on the formalized description of traffic spatio-temporal behavior from both packet level and flow level. In this model, we employ Convolutional Neural Network (CNN) to extract spatial information and Long Short-Term Memory (LSTM) to extract temporal information. In addition, we present a dataset consisting of Benign and Trojan HTTP Traffic (BTHT-2018). Experimental results show that our model can guarantee high accuracy (the F1 of 98.62\%$ \sim $99.81\% and the FPR of 0.34\%$ \sim $0.02\% in BTHT-2018). More importantly, our model has a huge advantage over other related methods in generalization ability. HSTF-Model trained with BTHT-2018 can reach the F1 of 93.51\% on the public dataset ISCX-2012, which is 20+\% better than the best of related machine learning methods.
		\end{abstract}
		
		
	\end{frontmatter}
	
	
	\section{Introduction}
	\label{sec1} 
	 Trojan, especially HTTP-based Trojan, is used by attackers to pass control instructions and perform malicious actions. Nowadays, on average, nearly 1.74 million host IP addresses in the global Internet were infected with "Flying" worm Trojan every month\cite{Xbookhulianwangchina}. With HTTP traffic as the carrier, attackers use Trojan scripts to transmit malicious information in the network, which brings great threats to network security of equipment and data. Moreover, it is difficult to distinguish the HTTP-based Trojan traffic from benign traffic because of its concealment and confusion.
	
	The network equipment and data security issues caused by HTTP-based Trojan have attracted more attention from people. HTTP-based Trojan traffic detection belongs to intrusion detection. Generally, people deploy intrusion detection systems (IDSs) to prevent network attacks. Anomaly detection is one of the main methods for intrusion detection field. It can detect unknown (0-day) attacks by constructing effective feature engineering of malicious and benign behaviors\cite{liao2013intrusion}. At present, researchers have proposed many anomaly detection methods based on machine learning (ML-based) and deep learning (DL-based), for detection of specific attacks (XSS\cite{gupta2017cross}, DDoS\cite{mirkovic2004taxonomy}, \textit{etc.}). There is less research on HTTP-based Trojan traffic detection. Moreover, these methods rely heavily on expert knowledge for the design of feature engineering and are limited by experimental scenarios. However, it is difficult for the original detection method or feature engineering to maintain excellent detection performance under new conditions\cite{hubballi2014false}. designing feature engineering requires expert knowledge and designing effective feature engineering for a specific application scenario is a hard work\cite{zhang2013effective}. In some cases, it is not even possible.
	
	Deep learning (DL) is widely applied and researched because of its powerful feature learning capabilities over the past several years. In intrusion detection, researchers propose DL-based methods to detect various specific network attacks (malicious traffic\cite{kim2017method, yin2017deep},  malware\cite{wang2019effective}, \textit{etc.}). After inputting low-dimensional features, DL can automatically abstract them into high-dimensional features through hierarchical transfer. 
	
	According to survey, there is currently no special HTTP-based Trojan traffic dataset. Although there is related malicious traffic in some well-known datasets\cite{ring2019survey, shiravi2012toward, tavallaee2009detailed}, it is relatively small and outdated for HTTP-based Trojan traffic, so that it is difficult to cover the full picture of HTTP-based Trojan attack patterns. Therefore, we provide a real dataset consisting of Benign and Trojan HTTP Traffic (BTHT-2018) to support related research.
	
	In this paper, we build a detection model based on DL knowledge. Based on the feature analysis of significant real traffic data, our model can learn the essential characteristics better, and obtain stronger interpret ability and generalization. In general, the contributions of this paper are as follows. 
	
	\begin{itemize}
		\item For HTTP-based Trojan detection, we present a formalized description of traffic spatio-temporal behavior from both packet level and flow level. Based on temporal and spatial dimensions, we take the Trojan's sending behavior (on-line, heartbeat, \textit{etc.}) and receiving behavior (control instruction, \textit{etc.}) as key monitoring steps to monitor traffic at the gateway. Traffic statistics characteristics are extracted from packet level and flow level, and then data is preprocessed and encoded by feature encoders.
		\item HSTF-Model (Model based on Hierarchical Spatio-Temporal traffic Features), a DL-based hierarchical hybrid structure neural network, is proposed, based on characteristics of the data on two attributes of temporal and spatial. In the temporal domain, a flow usually consists of multiple packets, and there is a timing relationship between packets. We use LSTM to extract their temporal characteristics. In the spatial domain, most flow payloads are large-scale, with both structural and textual features. Therefore, we use CNN to accelerate the convergence speed while extracting features. Also, we use fully-connected Multilayer Perceptron (MLP)\cite{rosenblatt1958perceptron} to process statistics and maximize its feature attributes. Experiments show that the model has excellent robustness and generalization. In the experiment, HSTF-Model can obtain the F1 of 99.47\% when detecting HTTP-based Trojan traffic.
		\item We built a prototype system based on HSTF-Model. Experimental verification was performed in dataset BTHT-2018 and public dataset ISCX-2012 that we collected and cleaned. In robustness, the model can reach F1 of 98.62\% in imbalanced dataset with the ratio of 1 : 100 (malicious : benign). In generalization, models trained with BTHT-2018 can achieve 91.14\% precision, 95.72\% recall, and 93.51\% F1 on ISCX-2012, while other methods can only reach 73.5\% F1 at most. In addition, there is currently no special HTTP-based Trojan traffic dataset. Therefore, we provide BTHT-2018 to help researchers to further investigate such attacks \footnote{The dataset can be found at \href{$https://github.com/ComputersSecurity/BTHT-2018$}{https://github.com/ComputersSecurity/BTHT-2018} and \href{$https://drive.google.com/open?id=1d\_SVIOzzgw2kYPlC5dKjgOl51YXTDZUi$}{https://drive.google.com/open?id=1d\_SVIOzzgw2kYPlC5dKjgOl51YXTDZUi}. Researchers who are going to use the dataset should indicate the original source of data by citing this paper.}. 
	\end{itemize}
	
	The remainder of this paper is organized as follows. HTTP-based Trojan scene analysis is described in Section 2, and the introduction of traffic data structure and feature in Section 3. Section 4 introduces the methodology of the model. We conduct experiments in Section 5. Related works are described in Section 6, Subsequently, we discuss the model and experiments in Section 7. Section 8 draws our conclusion and gives future research.
	
	\section{HTTP-based Trojan scene analysis}
	
\begin{figure*}[htbp]
		\centerline{\includegraphics[scale=0.45]{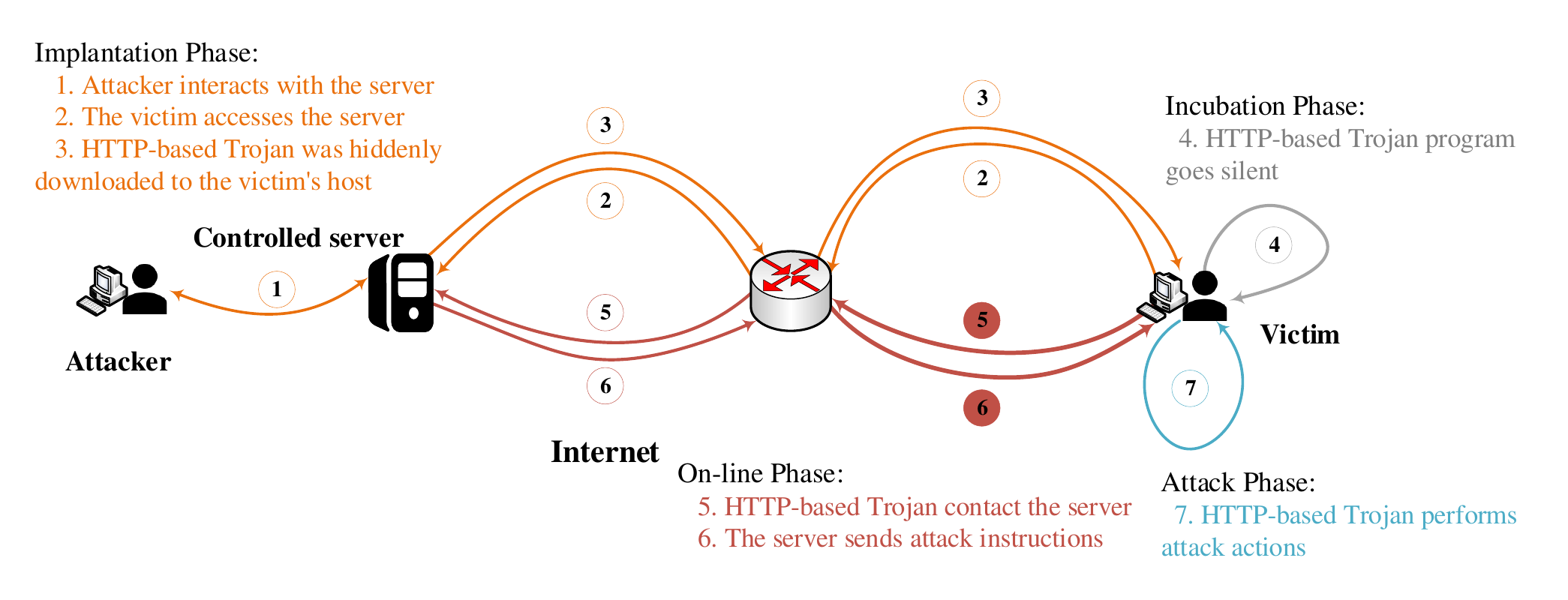}}
		\vspace{-1.3em}
		\caption{General attack process of HTTP-based Trojan.}
		\label{Trojans}
		\vspace{-0.5em}
	\end{figure*}

	In this paper, we show the general process of HTTP-based Trojan deployment and implementation of attacks, as shown in Fig.~\ref{Trojans}. It can be divided into 4 phases: implantation phase, incubation phase, on-line phase, and attack phase.
	
	During the implantation phase, the attacker uploads to the controlled web server. Then, the script is downloaded to the host when the victim host accesses the server. Then, the HTTP-based Trojan script enters the incubation phase. After successfully lurking, it will contact the controlled server from time to time, such as sending heartbeat packets. At this time, the script enters the on-line phase. After receiving the HTTP-based Trojan script's contact information, the attacker will send instructions to remotely control the HTTP-based Trojan to perform malicious actions (stealing private information, further intrusions, deploying springboards, \textit{etc.}). The HTTP-based Trojan script successfully received the instruction means that it entered the attack phase and began to carry out the attack. 
	
	The operation of the victim is unpredictable due to human factors, so the HTTP-based Trojan is difficult to prevent during the implantation phase. During the incubation phase, the device and information are still safe, and the HTTP-based Trojan will not show malicious actions, which makes it difficult for us to discover the insecure factors. If the Trojan is discovered during the attack phase, it is too late, at which point it has started executing malicious instructions in the victim's host.

	During the on-line phase, the HTTP-based Trojan starts to communicate with the outside world, sending on-line data packets and receiving instructions, which have spatio-temporal characteristics. In spatial, the HTTP-based Trojan generates packet data that is different from benign network behavior. The character distribution in the packet and the relative position between the characters (like the pixels of the picture) are distinguishable. In temporal, multiple packets of a flow have a natural temporal relationship. The flow generated by the HTTP-based Trojan and benign network behavior has different packet sequence characteristics (size, number, \textit{etc.}), which is also effective discrimination for detecting HTTP-based Trojan and benign traffic. The corresponding formal description is shown below.
	
	During the on-line phase, a HTTP-based Trojan generates data stream $ bflow = (bpkt_1, $ $bpkt_2, ..., bpkt_N) $ , the benign network behavior produces data stream $ wflow = (wpkt_1, wpkt_2, $ $..., wpkt_M) $.
		
	Let there be a spatial information discrimination function $ f_{spatial}(pkt1, pkt2) = \lambda (0 \leq  \lambda \leq 1)$ of packets. When $ pkt1=pkt2 $, $ \lambda = 0 $, when $ pkt1 \neq  pkt2 $, $ \lambda > 0 $. Then for $ bflow $ and $ wflow $, there is:
		
	\begin{equation*}
		\begin{aligned}
		f_{spatial}(bpkt_i, wpkt_j) > 0 \ \left ( i=1,2,...,N; j=1,2,...M\right )
		\label{eqd1}
		\end{aligned}
	\end{equation*}
		
	Let there be a temporal information discrimination function $ f_{temporal}(flow1, flow2) = \lambda (0 \leq  \lambda \leq 1)$ of packets. When $ flow1=flow2 $, $ \lambda = 0 $, when $ flow1 \neq  flow2 $, $ \lambda > 0 $. Then for $ bflow $ and $ wflow $, there is:
		
	\begin{equation*}
		\begin{aligned}
		f_{temporal}(bflow, wflow) > 0 
		\label{eqd2}
		\end{aligned}
	\end{equation*}
	
	This paper focuses on HTTP-based Trojan malicious behavior. When the attack enters on-line phase to communicate with the outside world, data packets it sends and receives usually have different characteristics from benign data based on spatio-temporal characteristics. Under the premise of ensuring the security of equipment and data, we can effectively detect the malicious behavior of the HTTP-based Trojan.

	\section{Traffic data structure and feature analysis}
	In this paper, a sample is a flow. A flow is defined as packets traveling between two computer addresses using a particular protocol on a particular pair of ports\cite{moore2013discriminators}. Packets with the same tuple of information ($ host_{src} $, $ port_{src} $, $ host_{dst} $, $ port_{dst} $, $ HTTP $) belong to the same flow. We define that flow is full-duplex. That is, the request and response data belong to the same flow.
	
	\subsection{\textbf{Traffic data structure}}
	We collected data from the actual network environment and generated a dataset BTHT-2018 (Benign and Trojan HTTP Traffic) after data cleaning and statistical analysis. Also, we extracted HTTP-based Trojan related traffic from ISCX-2012 and processed these into the same data format as BTHT-2018. For details, see section 5.2.

	BTHT-2018 consists of BTHT-R (Raw) and BTHT-S (Statistical), which represent the raw traffic data and the corresponding statistical features. The data types include HTTP-based Trojan traffic and Benign traffic. The former comes from network operators. and the latter is collected at a laboratory gateway. 
	
	\begin{figure}[htbp]
		\vspace{-0.5em}
		\centerline{\includegraphics[scale=0.8]{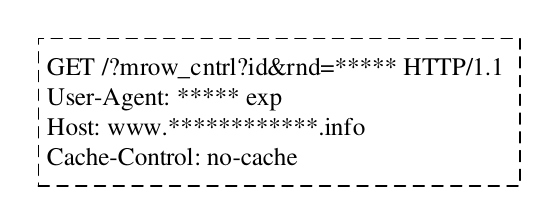}}
		\vspace{-1.5em}
		\caption{On-line packet generated by a HTTP-based Trojan script.}
		\label{On-line}
		\vspace{-0em}
	\end{figure}
	
	Fig.~\ref{On-line} shows the on-line package of a HTTP-based Trojan. It can be seen that it has features on key fields of structure and text. Because of the data security protection, the sensitive data (Host, IP, \textit{etc.}) of these flows are irreversibly hashed. In order to ensure data consistency, benign traffic uses the same processing technology.
	
	\subsection{\textbf{Statistical feature analysis}}
	Raw traffic data has valuable statistical features. We use expert knowledge to combine statistics and raw information, and then, extract comprehensive features based on deep learning. The url length of malicious traffic, for instance, is usually longer than that of benign urls. These statistical features are useful in identifying the type of traffic. In this paper, a raw flow sample generates a corresponding statistical feature sample. The raw dataset is dataset BTHT-R, and these statistical feature samples constitute the corresponding dataset BTHT-S.	
	
\begin{figure*}[htbp]
		\vspace{-1.5em}
		\centerline{\includegraphics[scale=0.8]{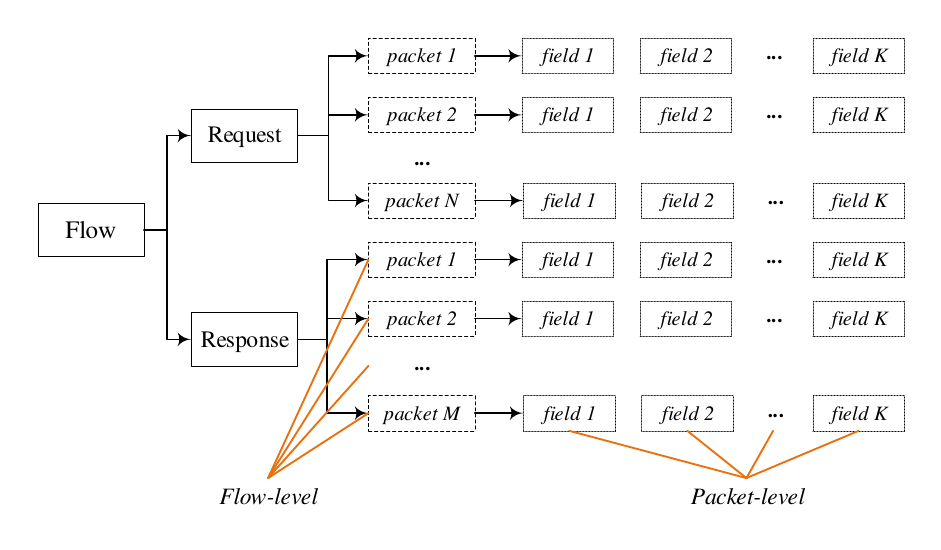}}
		\vspace{-1.5em}
		\caption{Hierarchical structure of HTTP traffic data.}
		\label{Hierarchy of a flow}
		\vspace{-0.5em}
	\end{figure*}

	Traffic is just like the written content produced by people communicating with language. It consists of paragraphs, sentences, phrases and words. Similarly, an HTTP flow is also hierarchical. Fig.\ref{Hierarchy of a flow} illustrates this feature. An HTTP flow consists of multiple packets. A packet consists of a header line, multiple field lines, and a payload. Therefore, we also use this property to build hierarchical statistical features.
	
	We extract features from two levels to form two feature vectors, packet-level (PL) and flow-level (FL), which take advantage of the hierarchical temporal nature of HTTP traffic data. Neural networks can further improve detection performance based on these statistical features.
	
	\subsubsection{\textbf{Packet-level statistical feature}}
	The packets of HTTP traffic can be divided into requests and responses. The statistical features of the request packets and response packets constructed in this paper are shown in Tab.~\ref{PL-vector}. A request packet mainly counts the request method, url length, field name and field value length of each field row, and payload length. A response packet mainly counts the response mode and the length of the status description. RFC1998\cite{chen1996rfc1998} recommends using 47 field lines for HTTP-based web services, but most web services do not use that much. In our dataset, 99.9\% of the data uses only a few field features. Therefore, we choose to count the data of 18 fields. Surely, this value can be adjusted according to different application scenarios. We use the method of replacing superfluous fields and deficient fields with 0 to maintain consistent data. In this paper, a packet generates a $ 1 \times 41 $ vector.
	
	\begin{table}[htbp]
		\vspace{-1em}
		\caption{Composition of the packet-level vector (PL-vector).}
		\vspace{-0.5em}
		\begin{center}
			\begin{tabular}{|c|c|c|c|}
				\hline
				\cline{1-4}
				\multicolumn{2}{|c|}{\textbf{Request}} &\multicolumn{2}{|c|}{\textbf{Response}} \\
				\hline
				Item Type & Position & Item Type & Position \\
				\hline
				\hline
				
				Req type & 0 & Res type & 0\\
				Length of url & 1 & Length of state description & 1\\
				Protocol version & 2 & Protocol version & 2\\
				Lines of fields & 3 & Lines of fields & 3\\
				Length of fields name & 4-21 & Length of fields name & 4-21\\
				Length of fields value & 22-39 & Length of fields value & 22-39\\
				Length of payload & 40 & Length of payload & 40\\
				\hline
			\end{tabular}
			\label{PL-vector}
		\end{center}
	\vspace{-2em}
	\end{table}
	
	\subsubsection{\textbf{Flow-level statistical feature}}
	A flow consists of multiple request packets and response packets, as a paragraph consists of multiple sentences. We usually analyze a paragraph through multiple sentences. Similarly, in the network, characteristics of traffic at the flow level can reflect more behavioral information and show the attacker's intention more comprehensively. In this paper, we divide the flow into two flow-level sequences, request sequence and response sequence, and then extract FL-vector separately.
	
	The first is the request sequence. We count the number of common packets, the number of request types, and the byte size sequence. The second is the response sequence. We count the number of packets, the number of response types, and the byte size sequence. We combine the analysis of the dataset to specify that a sequence will not generate more than 50 packets. If it exceeds, it will be discarded. If it is missing, it will be filled with 0. Of course, this can also be adjusted flexibly. The composition format of the statistical feature of request request and response sequence is shown in Tab.\ref{FL-vector}. In this paper, a request sequence constitutes a $ 1 \times 57 $ feature vector, and a response sequence generates a $ 1 \times 58 $ feature vector.
	
	\begin{table}[htbp]
		\vspace{-1em}
		\caption{Composition of the flow-level vector (FL-vector).}
		\vspace{-0.5em}
		\begin{center}
			\begin{tabular}{|c|c|c|c|}
				\hline
				\cline{1-4}
				\multicolumn{2}{|c|}{\textbf{Request}} &\multicolumn{2}{|c|}{\textbf{Response}} \\
				\hline
				Item Type & Position & Item Type & Position \\
				\hline
				\hline
				
				Count of req pkts & 0 & Count of res pkts & 0\\
				Count of 'get' & 1 & Count of '1XX' & 1\\
				Count of 'post' & 2 & Count of '2XX' & 2\\
				Count of 'head' & 3 & Count of '3XX' & 3\\
				Count of 'options' & 4 & Count of '4XX' & 4\\
				Count of other requests & 5 & Count of '5XX' & 5\\
				Mean of pkt bytes & 6 & Count of other responses & 6 \\
				Seq of pkts bytes & 7-56 & Mean of pkt bytes & 7 \\
				& & Seq of pkts bytes & 8-57 \\
				\hline
			\end{tabular}
			\label{FL-vector}
		\end{center}
	\vspace{-1.5em}
	\end{table}
	
	An raw flow generates PL-vector and FL-vector through feature analysis. The two become a sample in BTHT-S, which will be used as supplement to the raw data in subsequent experiments for traffic behavior analysis. Feature analysis uses empirical expert knowledge, which can provide more comprehensive feature information for model discrimination.
	
	\section{Model methodology}
	In this section, we build a hybrid structure neural network model based on deep learning, called HSTF-Model. We introduce the structure of the model and the main network components, and analyze the complexity of the model.
	
	\begin{figure*}[htbp]
		\vspace{-1.5em}
		\centerline{\includegraphics[scale=1.0]{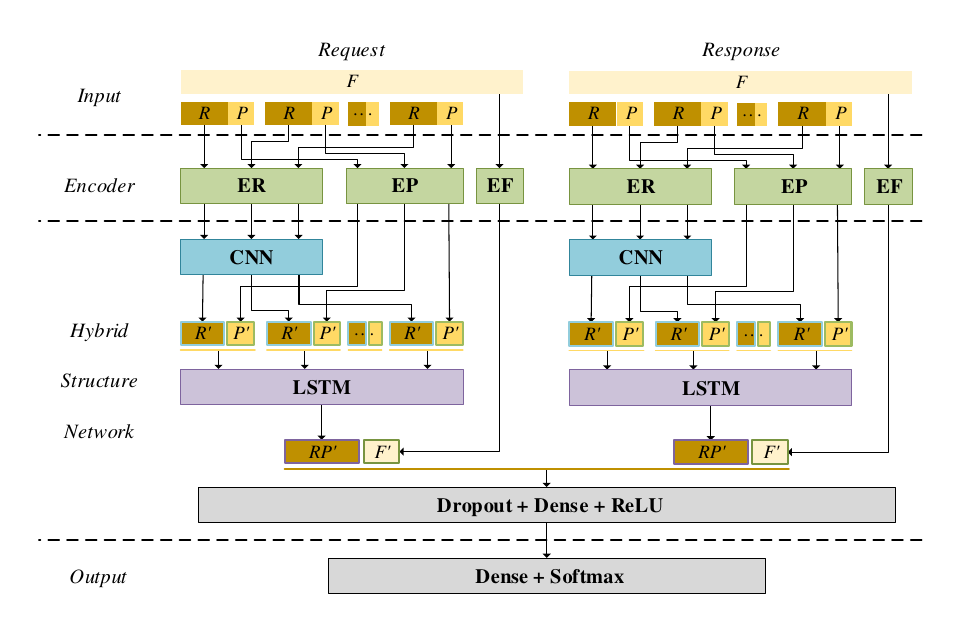}}
		\vspace{-1.5em}
		\caption{Architecture of HSTF-Model. (\textbf{F}: FL-vector; \textbf{P}: PL-vector; \textbf{R}: Raw-data; \textbf{$ \textbf{*}' $}: Intermediate variables ($ R' $, \textit{etc.}); \textbf{EF}: Feature Encoder of statistical data at flow-level; \textbf{EP}: Feature Encoder of statistical data at packet-level; \textbf{ER}: Feature Encoder of raw data).}
		\label{HSTF-Model}
		\vspace{-1em}
	\end{figure*}
	
	\subsection{\textbf{Overview}}
	HSTF-Model model structure is shown in Fig.~\ref{HSTF-Model}. A flow is divided into request and response. In addition to the raw data, there are a series of spatio-temporal sequence statistical features. Feature encoders based on MLP are used for feature encoding. Subsequently, CNN is designed to process the raw packet feature to extract spatial information. The raw packet features are combined with the corresponding statistical features to form feature vectors at the packet level. Multiple feature vectors form a temporal feature sequence, and the aggregated features are extracted by the LSTM. Finally, the processing results are combined with statistical features at the flow level, and the discrimination results are output through the fully connected network. Experiments show that this combination of raw information and statistical features can effectively improve the performance of the model. And we can change the parameters of the model in the combination stage such as CNN output and the corresponding packet-level statistical feature size to indicate the credibility of the raw traffic and statistical features. 
	
	\subsection{\textbf{Encoder}}
	The encoder converts the input data into feature codes suitable for neural network processing. We use MLP to build multiple encoders. MLP is a kind of feedforward artificial neural network model. It maps each data input to a single output. The structure includes input layer, hidden layer, and output layer. Non-linear activation functions are used between layers, which allows MLP to handle non-linear problems (xor problems, \textit{etc.}). MLP is inherited and developed by DNN, adding more hidden layers and richer activation function methods to improve the ability of the model to fit complex functions. Therefore, MLP can be considered as simple DNN, and DNN is complex upgraded MLP.
	
	\begin{figure}[htbp]
		\vspace{-1em}
		\centerline{\includegraphics[scale=0.55]{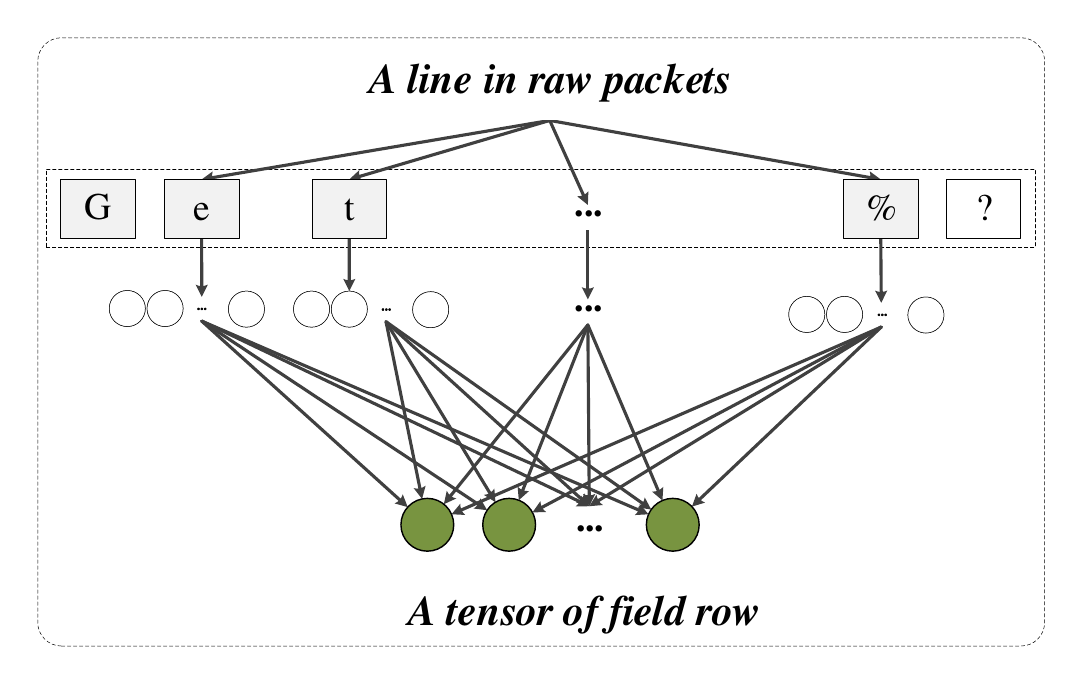}}
		\vspace{-1em}
		\caption{Feature encoder of raw data.}
		\label{ER}
		\vspace{-0.5em}
	\end{figure}
	
	For the raw packet data, the model extracts features from the traffic in the preprocessing stage and converts them into a standard format that can be processed by the neural network. The basic data preprocessing is to transform the packet into two-dimensional data in the form of an image, and each pixel is a text character tensor. The packets composed of multiple messages form a matrix sequence and enter the subsequent neural network through the feature encoder. The raw data feature encoder is shown in Fig.~\ref{ER}. The raw text after preliminary processing is transformed into a richer and suitable tensor representation through MLP.
	
	Before entering the encoder, we normalized the data to generalize the distribution of the unified sample to speed up the convergence of the model. Since most of the HTTP data is ascll encoded, we directly use the mod (\%) 128 operation to scale a character's ascll value to between $ 0\sim 1 $. In this way, the data is transformed as a whole, and the original relative relationship between the data can be maintained while the values are normalized to a specific interval. The preprocessing algorithm for text data is described in Algorithm~\ref{alg1}.
	
\begin{figure}[htbp]
		\vspace{-1em}
		\centerline{\includegraphics[scale=0.6]{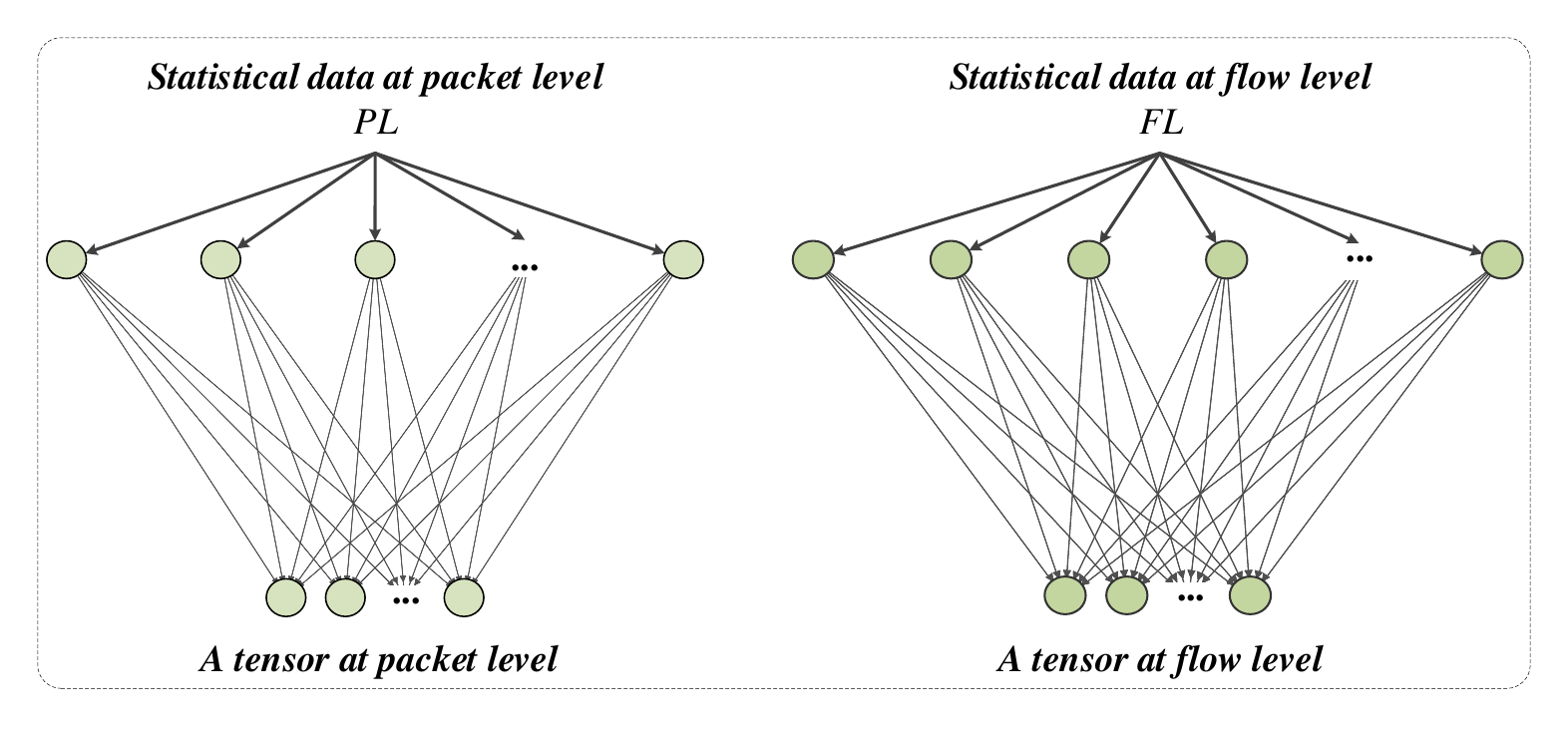}}
		\vspace{-1.5em}
		\caption{Feature encoder of statistical data at packet level and flow level.}
		\label{EPEF}
		\vspace{-0.5em}
	\end{figure}

	For the spatio-temporal sequence statistical features, at the packet level, a request packet and a response packet each generate a $ 1 \times 41 $ statistical vector. At the flow level, the request sequence and response sequence generate $ 1 \times 57 $ and $ 1 \times 58 $ feature vectors, respectively. A flow sample generates four statistical vectors of $ n \times 41 $, $ m \times 41 $, $ 1 \times 57 $, and $ 1 \times 58 $, where $ n $ and $ m $ represent the number of request and response packets in the flow. There is no clear correlation between statistical features, the amount of data input is small, and feature relationships are relatively easy to extract. Therefore, we built encoders based on MLP to process the statistical features, learn the statistical features, derive more data expressions, and further transform the feature vectors into multiple feature combination representations. As shown in Fig.~\ref{EPEF}, the statistical feature encoders are divided into PL feature encoders and FL feature encoders, but the difference is only in the difference of neurons, so they have the same preprocessing process.
	
	Before entering the feature encoder, we also used normalization to scale it to $ 0 \sim  1 $ on the basis of maintaining the relative relationship of the statistical data, which is more convenient for learning by the encoder. As a rule of thumb, we propose a sigmoid-like function as the normalization method for statistical data\cite{mira1995natural}. The corresponding statistical feature processing algorithm is described in Algorithm~\ref{alg2}.
	
	\begin{algorithm} 
		\caption{Feature coding algorithm of raw data} 
		\label{alg1}  
		\begin{algorithmic}[1] 
			\REQUIRE $ raw\_pkt $: a raw packet. $ m $: matrix size. $ s1, s2, ..., sn $: Number of neurons in each layer of encoder.
			\ENSURE $ Image\_pkt $: Feature image matrix of raw packet.
			\STATE{Step 1: Convert the packet into standard matrix of numerical form}
			\STATE{ $\ \ \ raw\_pkt  = (int)raw\_pkt\ \%\ 128$}
			\STATE{ $ \ \ \ mat  = a\ matrix\ of\ m[0] \times m[1] $}
			\STATE{ $ \ \ \ mat[0]  = the\ first\ m[1]\ elements\ of\ raw\_pkt[0] $}
			\FOR{$i=1$ to $m[0]-2$} 
			\STATE{ $ mat[i]  = the\ first\ m[1]\ elements\ of\ raw\_pkt[i] $}
			\ENDFOR 
			\STATE{ $ \ \ \ mat[m[0]-1]  = the\ first\ m[1]\ elements\ of\ raw\_pkt payload $}
			
			\STATE{Step 2: Build n-layer feature encoder}
			\FOR{$i=1$ to $n$} 
			\STATE{ Layer-i = The full connectivity layer of $ si $ neurons}
			\ENDFOR 
			\STATE {$ \ \ \ $ raw\_encoder = Layer-1 + Layer-2 + … + Layer-n} 
			
			\STATE Step3: Compute feature images
			\STATE {$ \ \ \ Image\_pkt = raw\_encode(mat) $ } 
			
		\end{algorithmic} 
	\end{algorithm}
	
	\begin{algorithm} 
		\caption{Feature coding algorithm of statistical data} 
		\label{alg2} 
		\begin{algorithmic}[1]
			\REQUIRE $ stat\_seq $: a statistical sequence. $ s1, s2, ..., sn $: Number of neurons in each layer of encoder.
			\ENSURE $ Feature\_seq $: Statistical feature coding sequence.
			\STATE{Step 1: Normalize each element of statistical sequence}
			\STATE{ $\ \ \ stat\_seq = \frac{1-e^{-stat\_seq}}{1+e^{-stat\_seq}}$}
			
			\STATE{Step 2: Build n-layer feature encoder}
			\FOR{$i=1$ to $n$} 
			\STATE{ Layer-i = The full connectivity layer of $ si $ neurons}
			\ENDFOR 
			\STATE {$ \ \ \ $ stat\_encoder = Layer-1 + Layer-2 + … + Layer-n} 
			
			\STATE Step3: Compute feature images
			\STATE {$ \ \ \ Feature\_seq = stat\_encode(stat\_seq) $ } 
			
		\end{algorithmic} 
	\end{algorithm}
	
	\subsection{\textbf{CNN}}

	CNN belongs to deep feed-forward neural network that includes convolution and pooling operations. It has the capability of representation learning and can perform translation-invariant classification in the input data according to its hierarchical structure. The iterative upgrade of the computing power of the hardware equipment enables CNN to train faster. In addition, the hidden layer of CNN has the sparseness of convolution kernel parameter sharing and inter-layer connection, which makes it possible to effectively extract the spatial structure features of the data with a small amount of calculation without feature engineering.
	
	The core of CNN is the convolution and pooling layer, and the corresponding operations used in this paper are shown in Eq~\eqref{eq1} and Eq~\eqref{eq2}. The cross-correlation calculation is performed in the output feature map $ Z^l $ of the $ l $ layer and the convolution kernel $ w^{l + 1} $ of the $ l + 1 $ layer. The output of the convolutional layer is formed, and then the pooling operation (if it exists) is performed to form the convolutional overall output of the $ l + 1 $ layer. $ K $ is the number of channels of the feature map, $ f $ is the corresponding convolution kernel and pooling size, $ s_0 $ is the stride, and $ p $ is the number of padding. For structured information, each convolution kernel traverses the input features regularly, performs matrix multiplication and superimposes the amount of deviation in the features. Multiple convolution kernels can extract high-order global features of the data from multiple angles. The function of the pooling operation is feature selection and information filtering, and processing downsampling of the convolution kernel output. According to the preset pooling function, the pooling layer can convert point features into regional features, further aggregate features and reduce model overfitting.
	
	\begin{equation}
	\begin{aligned}
	Z^{l+1}(i,j)&=[Z^{l}\otimes w^{l+1} ](i,j)+b \\
	&= \sum_{k=1}^{K_l}\sum_{x=1,y=1}^{f}\left [ Z_{k}^{l} (s_0i+x,+s_0j+y)w_{k}^{l+1}(x,y))\right ]+b 
	\label{eq1}
	\end{aligned}
	\end{equation}
	
	\begin{equation}
	\begin{aligned}
	Z_{k}^{l}(i,j)&=\left [ \sum_{x=1,y=1}^{f} Z_{k}^{l}(s_0i+x, s_0j+y)^{p}\right ]^{\frac{1}{p}}
	\label{eq2}
	\end{aligned}
	\end{equation}
	
	Based on the structural characteristics of the packet and the advantages of the CNN, we use CNN to process the raw packet data. A packet is preprocessed to form a two-dimensional 'graphic' with a channel 1 similar to an image. We extract pixel features based on convolution kernel, a single convolution kernel extracts local features, and multiple convolution kernels form a convolution layer to extract the overall features. Then, we use maximum pooling ($ p \to +\infty $ in Eq~\eqref{eq2}) to perform feature filtering and processing in the convolution layer output. The activation function $ ReLU $ is connected between network layers as an activation function to inject nonlinear features into the model.
	
	\subsection{\textbf{LSTM}}

	LSTM is an improvement of Recurrent Neural Network (RNN). Traditional RNNs have the disadvantages of memory loss and the inability to control the nonlinear relationship of long-span sequence information. When the sequence is too long, traditional RNNs will experience gradient disappearance or explosion, which causes it to use only neighboring data features. LSTM uses gate structure to selectively remember and forget past information. A gate consists of an activation layer and a pointwise operation. By learning the previous data and using it in the current task, the gate structure allows the LSTM to selectively store information for subsequent processing, which allows information to be passed further along the timing chain. 
	
	Sundermeyer \emph{et al.} \cite{sundermeyer2012lstm} show that the most important structure in the LSTM is the Forget gate, followed by the Input gate, and finally the Output gate. Three gate structures constitute an LSTM cell, an intelligent network unit. We enter this cell through loop input sequence data, which can record the long-term dependent data characteristics.
	
	Before storing new information, cell will choose to discard a part of the information passed in the previous sequence. The forget gate, $ f_t $, gives a value of 0-1 according to the stored information $ H_{t-1} $ and the input $ x_t $ at the current stage, and selects the proportion of the retained information.
	
	\begin{equation}
	\begin{aligned}
	f_{t}&=\delta (W_{f}\cdot \left [h_{t-1}, x_{t}\right ]+b_{f})
	\label{eq3}
	\end{aligned}
	\end{equation}
	
	After discarding the information, the input gate, $ i_t $, decides how much new information to add.
	
	\begin{equation}
	\begin{aligned}
	i_{t}&=\delta(W_{i}\cdot \left[h_{t-1}, x_{t}+b_{i}\right] )
	\label{eq4}
	\end{aligned}
	\end{equation}
	
	Based on the current task, cell decides the information $ \tilde{C_t} $ that needs to be updated. The input and output gates combine to determine the latest information $ C_{t} $.
	
	\begin{equation}
	\begin{aligned}
	\tilde{C_{t}}&=\tanh (W_{C}\cdot \left [h_{t-1}, x_{t} \right]+b_{C})\\
	C_{t}&=f_{t}\times C_{t-1}+i_{t}\times \tilde{C_{t}}
	\label{eq41}
	\end{aligned}
	\end{equation}
	
	After the new information is stored, we need to determine the output state $ h_t $ of the cell, which is controlled by the output gate, $ o_t $.
	
	\begin{equation}
	\begin{aligned}
	o_{t}&=\delta (W_{o}\cdot \left [h_{t-1}, x_{t}\right ]+b_{o}) \\
	h_{t}&=o_{t} \times \tanh(C_{t})
	\label{eq42}
	\end{aligned}
	\end{equation}
	
	The LSTM can automatically select the processed and transmitted information through these three gate structures, so that it can effectively avoid the problem of long-term dependence.
	
	In the flow, a packet is processed with CNN and MLP to output feature data of the same dimension. These data are naturally temporal sequence because of the conversation. The latter data depends on the previous data. And the dependencies in the data sequence may have a long span, due to some factors such as network delay and retransmission. Therefore, we use LSTM to process these feature data. After the previous feature data is processed by the LSTM cell, it is passed to the subsequent tasks. LSTM cell outputs the processing result of the whole sequence at the last moment. These features are combined with the statistical feature coding at the flow level to enter the subsequent network layer.
	
	\subsection{\textbf{Complexity}}
	We analyze the complexity of HSTF-Model. Under the premise of determining the data feature and model parameters, the training time complexity of the model is linear with the sample size, and the space complexity is constant. 
	
	\subsubsection{\textbf{Time complexity}}
	Time complexity is the number of statement executions in an algorithm, called statement-frequency or time-frequency. In this paper, the time complexity of neural networks is the number of operations the model performs in training.
	
	For CNN, the time complexity is the sum of the computing time of each layer. A single convolution kernel needs to traverse all the inputs, and the time complexity is the product of the size of the convolution kernel and the size of the input data. The overall time complexity of CNN is shown in Eq~\eqref{eq8}, $ D_{CNN} $ is the depth of CNN, $ M $ is the size of a convolution kernel output feature map, $ K $ is the convolution and size, $ C_{l-1} $ is the number of output channels of the previous layer,$ C_l $ is the number of output channels of the current layer, that is, the number of convolution kernels ($ C_0 = (P \cdot n_{req} \cdot n_{res}) $, $ n_{req} $ and $ n_{res} $ are the size of the request and response in a flow, and $ P $ is the packet size).
	
	\begin{equation}
	\begin{aligned}
	Time_{CNN}\sim O\left ( \sum_{l=1}^{D_{CNN}} M_{l}^{2}\cdot K_{l}^{2}\cdot C_{l-1} \cdot C_{l}\right )
	\label{eq8}
	\end{aligned}
	\end{equation}
	
	For LSTM, time is mainly consumed by the length of the input sequence, as shown in Eq~\eqref{eq9}, $ L $ is the sequence length ($ L = n_{req} + n_{res}$), and $ C_{LSTM} $ is the cell state dimension.
	
	\begin{equation}
	\begin{aligned}
	Time_{LSTM}\sim O\left ( C_{LSTM} \cdot L \right )
	\label{eq9}
	\end{aligned}
	\end{equation}
	
	For fully connected layers, neurons between layers are connected to each other, and the operations between layers are multiplicative. Therefore, the time complexity is shown in Eq~\eqref{eq10}, which is the multiplication of the neuron state dimensions of each layer, $ D_{FC} $ is the number of layers, and $ S_{l} $ is the number of neuron of $ l $ layer.
	
	\begin{equation}
	\begin{aligned}
	Time_{FC}\sim O\left ( \prod_{l=1}^{D_{FC}}S_{l} \right )
	\label{eq10}
	\end{aligned}
	\end{equation}
	
	The time complexity of our proposed HSTF-Model is mainly composed of sample size, CNN processing, LSTM processing, and fully connected layer, as shown in Eq~\eqref{eq11}, where $ N $ is the sample size.
	
	\begin{equation}
	\begin{aligned}
	Time_{HSTF-Model} &= N \cdot \left (Time_{CNN} + Time_{LSTM}+Time_{FC} \right )\\
	&\sim O\left(\alpha \cdot N\right) \sim O\left( N\right)
	\label{eq11}
	\end{aligned}
	\end{equation}
	
	In Eq~\eqref{eq11}, $ \alpha $ is a variable determined by the data feature size and model parameters, when these are determined, it is a constant. In general, for time complexity, there is a linear relationship with the sample size.
	
	\subsubsection{\textbf{Space complexity}}
	Space complexity is a measure of the storage space required for an algorithm to execute within a computer. The spatial complexity of the neural network is the sum of its total parameters and the output feature maps of each layer. The total parameter amount is the sum of the weight and bias parameters of each neuron in the network. The output feature map of each layer is the temporary space occupied by the output of the neuron when processing data.
	
	The space complexity of the parameter quantity of HSTF-Model is shown in Eq~\eqref{eq12}, which is mainly composed of CNN, LSTM, and fully connected layer, $ D_{CNN} $ is the depth of CNN, $ K $ is the convolution and size, $ C_{l-1} $ is the number of output channels of the previous layer, $ C_l $ is the number of output channels of the current layer. $ C_{LSTM} $ is the number of state of LSTM, and $ D_{FC} $ is the depth of fully connected layer, and $ S_{l} $ is the number of neuron states of $ l $ layer.
	
	\begin{equation}
	\begin{aligned}
	Space_{parameter} \sim O\left ( \left(\sum_{l=1}^{D_{CNN}} K_{l}^{2}\cdot C_{l-1} \cdot C_{l}\right) + C_{LSTM}+\left(\sum_{l=1}^{D_{FC}} S_{l}\right)\right)
	\label{eq12}
	\end{aligned}
	\end{equation}
	
	The output feature map complexity is the size of the feature map calculated by each network layer in the actual operation. After the depth parameter is fixed, space is mainly related to the amount of data. As shown in Eq~\eqref{eq13}, $ M $ is the output space size of CNN.
	
	\begin{equation}
	\begin{aligned}
	Space_{feature} \sim O\left ( \left(\sum_{l=1}^{D_{CNN}} M_{l}^{2} \cdot C_{l}\right) + C_{LSTM}\cdot n_{req}\cdot n_{res}+\left(\sum_{l=1}^{D_{FC}} S_{l}\right)\right)
	\label{eq13}
	\end{aligned}
	\end{equation}
	
	The overall spatial complexity of HSTF-Model is shown in Eq~\eqref{eq14}, $ b $ is the sample size for a batch.
	
	\begin{equation}
	\begin{aligned}
	Space_{HSTF-Model} &= Space_{parameter} + b \cdot Space_{feature} \\
	& \sim O\left(\beta \right) \sim O\left(1 \right)
	\label{eq14}
	\end{aligned}
	\end{equation}
	
	In Eq~\eqref{eq14}, $ \beta $ is determined by the batch data size and model size, when these are determined, it is a constant. In general, for space complexity is $ O(1) $.
	
	\subsubsection{\textbf{Comparison with other methods in complexity}}
	We analyze machine learning methods that are widely used in anomaly detection (SVM\cite{cortes1995support}, \textit{etc.}), and find the relationship between complexity and training sample size on the premise of determining sample features and model parameters. The results are shown in Tab~\ref{other-c}, $ N $ is the training sample size. The time and space complexity of HSTF-Model is the same as the optimal complexity. In the experiment in section 5.7, in addition, HSTF-Model has better detection and generalization performance.
	
	\begin{table}[htbp]
		\vspace{-1em}
		\caption{Complexity comparison of HSTF-Model with other methods.}
		\vspace{-0.5em}
		\begin{center}
			\begin{tabular}{|c|c|c|}
				\hline
				\multirow{2}{*}{\textbf{Method}} &\multicolumn{2}{|c|}{\textbf{Complexity}} \\
				\cline{2-3} 
				 &\textbf{\textit{Time}} & \textbf{\textit{Space}} \\
				\hline
				\hline
				Naive Bayes & $ O(N) $ & $ O(1) $\\
				\hline
				Decision Tree & $ O(NlogN) $ & $ O(1) $\\
				\hline
				SVM & $ O(N^2) $ & $ O(1) $ \\
				\hline
				HSTF-Model & $ O(N) $ &$ O(1) $\\
				\hline
			\end{tabular}
			\label{other-c}
		\end{center}
	\vspace{-1.5em}
	\end{table}
	
	\section{Experiment and evaluation}
	In this section, we build a prototype system and use different evaluation indicators and the same evaluation criteria to test the proposed detection method on real and public datasets. The optimal structure of the model was determined, the robustness and generalization were tested, and the results were analyzed and discussed.

\subsection{\textbf{Prototype system}}
We proposed a prototype system based on HSTF-Model, combined with traffic spatio-temporal behavior modeling, to effectively detect HTTP-based Trojan traffic. The algorithm is shown in Algorithm~\ref{alg3}. First, the data is preprocessed (normalized, \textit{etc.}) according to the sequence model, and transformed into tensors suitable for neural network processing. Subsequently, low-latitude features are input into HSTF-Model, and feature encoders are used to aggregate and unify the data features. The model extracts packet information through CNN, extracts sequence information through LSTM, to further learn and abstract features, and finally, discriminates and outputs detection results comprehensively.

\begin{algorithm} 
	\caption{Detection algorithm against malicious traffic } 
	\label{alg3} 
	\begin{algorithmic}[1] 
		\REQUIRE $ raw\_flow $ : a raw flow. $ \lambda $ : threshold to determine whether data is malicious. $ P $ : Relevant parameters of building model.
		\ENSURE $ class $: "Malicious" or "Benign".
		
		\STATE{Step 1: Preliminary processing data and Extract statistical features}
		\STATE{ $\ \ \ req\_pkt\_seq, req\_pl\_seq, req\_fl, res\_pkt\_seq, res\_pl\_seq, res\_fl = raw\_flow$}
		
		\STATE{Step 2: Build HSTF-Model}
		\STATE{$ \ \ \ $ model = HSTF-Model($ P $)}
		
		\STATE Step3: Compute results
		\STATE {p = model($ req\_pkt\_seq, req\_pl\_seq, req\_fl, res\_pkt\_seq, res\_pl\_seq, res\_fl $) } 
		\IF{$ p[1] $ $ > $$\lambda $} 
		\STATE class = "Malicious"
		\ELSE 
		\STATE class = "Benign" 
		\ENDIF 
	\end{algorithmic} 
\end{algorithm}

\subsection{\textbf{Data collection}}
We collected traffic from the actual network and generated a dataset BTHT-2018 to verify the detection method proposed in this paper. The data includes HTTP-based Trojan traffic and benign traffic, which are filtered and extracted through data cleaning technology (deletion of irrelevant data, flow reorganization, \textit{etc.}). At the same time, this paper analyzes and screens the public dataset ISCX-2012 to support the generalization verification of the model.

\subsubsection{\textbf{BTHT-2018}}
BTHT-2018 consists of BTHT-R and BTHT-S, which represent the raw traffic data and the corresponding spatio-temporal statistical features. Benign traffic accounts for 99\% of the dataset, which is derived from benign behavior, and HTTP-based Trojan traffic accounts for approximately 1\%, which is derived from malicious Trojans. The specific dataset details are shown in Tab~\ref{tab8}.

Benign traffic comes from the gateway exit of our network lab. After obtaining the authorization, we deployed a traffic collection device at the gateway under the premise of ensuring security and data privacy. Finally, we collected about 300GB of traffic data in a month, which is mainly divided into news browsing, social activities, web traffic, data download, and other types. Benign traffic is labeled based on trusted applications and trusted access object whitelist (IP, domain name, etc.). We filter the data to ensure that the data is benign traffic to the greatest extent. After removing error and irrelevant information through data cleaning, about 4 million benign traffic is obtained in units of flows.

HTTP-based Trojan traffic is generated by various Trojan horse attacks. The data of this paper is provided by network operators. There are two sources of malicious traffic data. One is based on malicious domain names, IP and other characteristic rules, using malicious traffic detection system to capture from the actual network. The other is to obtain malicious data by breeding malicious scripts under controlled and harmless conditions. In addition, we manually checked part of the marked malicious data to further enhance the correctness. After processing the data collected through these two methods, a total of about 37,000 flows were obtained. Data types include malicious downloads, stealing attacks, malicious promotion, and secondary implants of Trojans.

We performed irreversible desensitization on the published dataset to protect user privacy, while also eliminating the threat of malicious data and preventing it from damaging the network environment.

\begin{table}[htbp]
	\vspace{-1em}
	\caption{Statistics on packet size (in bytes) and flow size (in packets) in BTHT-2018.}
	\vspace{-1.5em}
	\begin{center}
		\resizebox{\textwidth}{15mm}{
		\begin{tabular}{|c|c|c|c|c|c|c|}
			\hline
			\multirow{2}{*}{\textbf{Statistics}} &\multicolumn{2}{|c|}{\textbf{Malicious}} &\multicolumn{2}{|c|}{\textbf{Benign}} &\multicolumn{2}{|c|}{\textbf{Total}} \\
			\cline{2-7} 
			& \textbf{\textit{Packet}} & \textbf{\textit{Flow}} & \textbf{\textit{Packet}} & \textbf{\textit{Flow}} & \textbf{\textit{Packet}} & \textbf{\textit{Flow}} \\
			\hline
			\hline
			Count & 2,842,054 & 37,847& 7,280,541& 4,044,741 & 10,122,595& 4,082,588 \\
			\hline
			Size &1,013,098,849&     2,842,054    & 4,319,581,422    & 7,280,541& 5,332,680,271&     10,122,595 \\
			\hline
			Min & 93&     1&     12    & 1&     12&     1 \\
			\hline
			Max & 8,444&     99&     10,427&     2,441&     10,427&     2,441 \\
			\hline
			Mean &356.47&     75.09    & 593.31    & 1.8    & 526.81    & 2.48\\
			\hline
			Var & 32.23&     39.06&     512.39&     7.53&     447.72&     10.94 \\
			\hline
		\end{tabular}}
		\label{tab8}
	\end{center}
	\vspace{-2em}
\end{table}

\subsubsection{\textbf{ISCX-2012}}

ISCX-2012 is a supervised network dataset in PCAP format published by the Canadian Institute for Cybersecurity of UNB in 2012, including actual traffic generated by HTTP, SMTP, SSH, IMAP, POP3, and FTP. By constructing various intrusion scenarios, the agency collected and tracked comprehensive interactive traffic for 7 consecutive days, and injected network attacks such as Infiltrating the network from inside, Distributed Denial of Service using an IRC Botnet within 4 days.

Based on the statistical analysis of the dataset ISCX-2012, we selected the traffic related to the Trojan and extracted benign and malicious HTTP flow data from it. Through the same processing method as BTHT-2018, 22,854 malicious samples and 222,494 benign samples were obtained. The specific dataset details are shown in Tab~\ref{tab9}. These data are combined with BTHT to verify the generalization performance of the model, that is, the detection effect of training on one dataset on another dataset.

\begin{table}[htbp]
	\vspace{-1em}
	\caption{Statistics on packet size (in bytes) and flow size (in packets) in ISCX-2012.}
	\vspace{-1.5em}
	\begin{center}
		\resizebox{\textwidth}{15mm}{
		\begin{tabular}{|c|c|c|c|c|c|c|}
			\hline
			\multirow{2}{*}{\textbf{Statistics}} &\multicolumn{2}{|c|}{\textbf{Malicious}} &\multicolumn{2}{|c|}{\textbf{Benign}} &\multicolumn{2}{|c|}{\textbf{Total}} \\
			\cline{2-7} 
			 & \textbf{\textit{Packet}} & \textbf{\textit{Flow}} &  \textbf{\textit{Packet}} & \textbf{\textit{Flow}} & \textbf{\textit{Packet}} & \textbf{\textit{Flow}} \\
			\hline
			\hline
			Count &  7,310,763 &  22,854& 2,224,730 & 222,494  & 9,535,493 & 245,348  \\
			\hline
			Size & 4,229,399,657&    7,310,763    &978,717,447    &2,224,730    &5,208,117,104    &9,535,493 \\
			\hline
			Min & 16&    1    &29    &1    &16&    1 \\
			\hline
			Max & 184,748&    813    &259,200    &3,531&    259,200&    3,531 \\
			\hline
			Mean & 578.52&    319.89    &439.93&    10    &546.18&    38.87 \\
			\hline
			Var &  454.76    &119.94    &662.57&    43.9&    514.21    &105.83 \\
			\hline
		\end{tabular}}
		\label{tab9}
	\end{center}
\vspace{-2em}
\end{table}
	
	\subsection{\textbf{Evaluation metrics and environment configuration}}
	\subsubsection{\textbf{Evaluation metrics}}
	Precision and recall are used as primary evaluation indicators to verify the detection performance of the model, as shown in Eq\eqref{eq5}. $ F_{\beta} $ is also calculated as a comprehensive evaluation index. A represents the weight of P and R in this index. The larger $ \beta (> 1)$ represents R is more important, and the smaller $ \beta (<1) $ represents P is more important. We set $ \beta = 1 $ to show that both are equally important.
	
	\begin{equation}
	\begin{aligned}
	Precision(P)&=\frac{TP}{TP+FP} \\ 
	Recall(R)&=\frac{TP}{TP+FN} \\
	F_{\beta }&=\frac{(1+\beta ^{2})\times P\times R}{(\beta ^{2}\times P)+R}
	\end{aligned}\label{eq5}
	\end{equation}
	
	 In addition, the false positive rate(FPR) and the true positive rate(TPR) (Eq~\eqref{eq6}) as the horizontal and vertical axes, respectively, are used for drawing ROC curves to show the expected generalization of the model. In all experiments in this paper, the ratio of malicious : benign in test set of  is always 1 : 1. Therefore, we can infer the value of FPR from P and R.
	 
	\begin{equation}
	\begin{aligned}
	FPR &=\frac{FP}{TN+FP}  = \frac{R \times (1-P)}{P} \\
	TPR &= \frac{TP}{TP+FN} = R\\
	\end{aligned}\label{eq6}
	\end{equation}
	
	\subsubsection{\textbf{Environment configuration}}
	HSTF-Model is implemented in Python3.5 based on the libraries of Keras and TensorFlow. The system environment of experiments is Ubuntu16.04 LTS. All software applications are deployed on a server machine with 64 CPU cores and 128GB memory. To further accelerate matrix computing, 3 NVIDIA TITAN XP are installed in the server.
	
	Based on experience and previous experimental basis, we determine some basic parameters of the model. The two local networks handling requests and responses have the same structure. The convolutional layer on CNN contains two $2\times 8$ convolution kernels with strides = 2. The maximum pooling layer size is $2\times 2$, and strides = 1. The hidden state dimension of the LSTM cell is 16. The model output dimension is 2, which represents the probability that the input data is malicious and benign. Moreover, we take dropout with 0.3 ratios behind LSTM to avoid overfitting, and the model is trained with $ Adam $ optimizer with learning rate = 0.0001.
	
	There are two sources of experimental data in this paper, one is the dataset BTHT-2018 collected and cleaned from the actual network, and the other is the public dataset ISCX-2012 obtained through processing. In the experiment, HSTF-Model not only uses the traffic in the same dataset for training and testing but also to verify the generalization ability of the model, we train the model on BTHT-2018 and apply it to ISCX-2012 to observe its detection ability. We also designed a variety of experimental scenarios with different proportions of malicious traffic and benign traffic to test the performance of the model in imbalanced datasets, that is, robustness. For each experiment, we randomly select data and repeat 10 times to take the average. There is no intersection between the train set and the test set. In addition, the ratio of malicious : benign in test set is always 1 : 1.
	
	\subsection{\textbf{Determination of key parameters}}
	The feature distribution of different datasets is different. Similarly, the flows are generally different. A flow may have many interactive packets, each of which is large, while the other flow may have only one packet and the size is very small. The difference between the two flows is obvious. In model detection, we need to determine the number and size of packets in the flow, which needs to be determined experimentally. However, the combination of the number and size of the packets is diverse, so we have adopted a method of controlling variables for determination. Because the request and response are equally important, we set the size of the request packet and response packet to the same size, and the number of corresponding packets is also set to the same size.
	
	\subsubsection{\textbf{Determination of optimal packet size}}
	Combined with the statistical analysis of the dataset, we set the number of packets, flow size = 3, in the experiment, and set the packet size = 100, 200, 300, 400, 500, 600, 800, 1000, 1200, 1600, 2400, 4000, 5000 bytes respectively. According to the experimental results, we determine the optimal packet size.
	
	\begin{figure}[htbp]
		\vspace{-1em}
		\centerline{\includegraphics[scale=1]{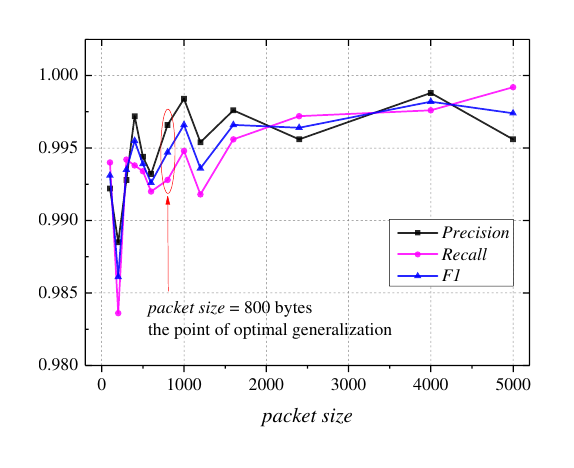}}
		\vspace{-2em}
		\caption{Effect of packet size on HSTF-Model (flow size = 3 packets).}
		\label{packet-size}
		\vspace{-0em}
	\end{figure}
	
	After HSTF-Model is trained in the dataset BTHT-2018, the detection results in the test set are shown in Fig.~\ref{packet-size}. It shows that a sample can provide less feature information when the packet size is small, and the detection effect of the model is relatively poor. Then, increasing the packet size can provide more information, and the indicators of the model are more than 99\%. The packet size has a better detection ability within 300-5000 bytes, which indicates that the model can obtain good results with fewer features. But this also means that with the increase of feature information, the detection effect of the model is not increased. This is because we mainly study the traffic of the Trojan horse during the on-line phase. The packets in this traffic are usually not too large and only include simple communication and command transmission. Therefore, it can be effectively judged by the data model in the previous part of the packets. Here we set the packet size = 800 as our selected optimal value because the model can have good detection results and generalization at the same time.
	
	\subsubsection{\textbf{Determination of optimal flow size}}
	The sequence of packets in HSTF-Model is processed using LSTM, so we need to determine the flow size (number of packets). In the experiment, we set flow size = $(1,2,3,4,5,6,$  $8,16,24,32,50) $ respectively and choose the optimal value through experiments.
	
	\begin{figure}[htbp]
		\vspace{-1em}
		\centerline{\includegraphics[scale=1]{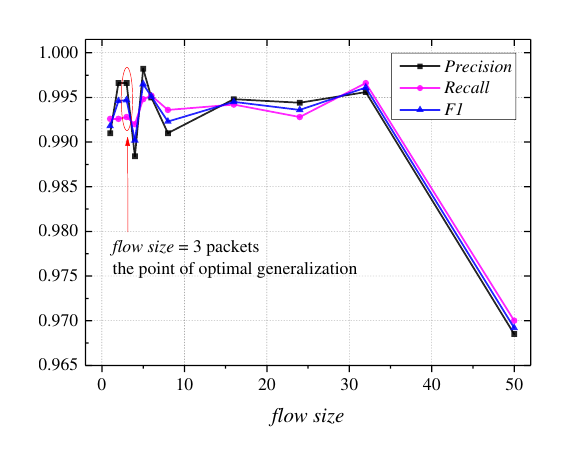}}
		\vspace{-2em}
		\caption{Effect of flow size on HSTF-Model (packet size = 800 bytes).}
		\label{flow-size}
		\vspace{-0.5em}
	\end{figure}
	
	Fig.~\ref{flow-size} shows the experimental results of determining the flow size. We choose flow size = 3 as the optimal value because the model can achieve the best balance between accuracy and generalization. During the on-line phase, the connection between the attacker and the HTTP-based Trojan basically does not include the transmission of a large amount of data, so the characteristics of the data are usually reflected in the first few packets of the flow. When the flow increases, the effective information contained in the newly added data will gradually decrease. When there is too much data, it will lead to a decrease in detection performance, because invalid data is equivalent to noise, which will interfere with the essential characteristics of the data extracted by the model.
	
	\subsection{\textbf{Efficiency of statistical features}}
	In order to evaluate the contribution of spatio-temporal sequence statistical features, we construct a contrast model (C-Model), which is a simplified HSTF-Model without extracting statistical features. The input only has raw data, the model only uses CNN to process the raw packet, and then LSTM extracts sequence features of CNN outputs. Finally, the model outputs the discrimination results through the fully connected layer.
	
\begin{figure*}[htbp]
		\vspace{-1.5em}
		\centerline{\includegraphics[scale=0.60]{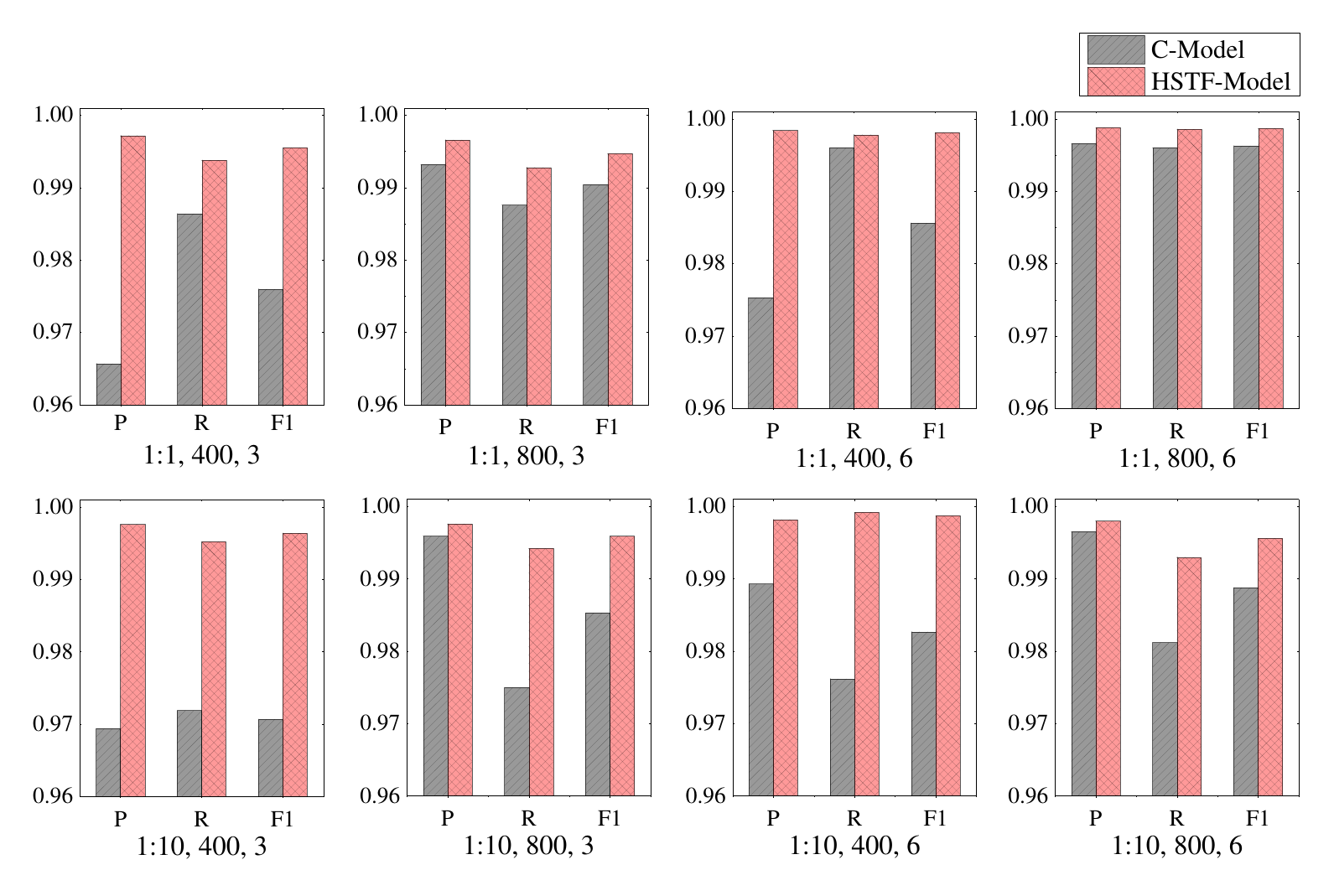}}
		\vspace{-1.2em}
		\caption{The detection performance of C-Model (the simplified HSTF-Model without extracting statistical features) and HSTF-Model in different scenarios. "1 : 1, 400, 3", for instance, means that the ratio of malicious : benign = 1 : 1 in training, packet size = 400 bytes and flow size = 3 packets.}
		\label{basic}
		\vspace{-0.5em}
	\end{figure*}

	We conducted experiments in scenarios with different proportions of malicious traffic in training sets, different packet sizes, and different flow sizes. The results are shown in Fig~\ref{basic}. Experiments show that HSTF-Model is superior to models without statistical feature sequences under various conditions. Because the combination of statistical features and raw traffic provides a richer representation for data samples, the model can relatively use fewer neurons to analyze and extract feature information, and it is easier to learn the essential features of the data.
	
	\subsection{\textbf{Robustess analysis of HSTF-Model}}
	
	\begin{figure}[htbp]
		\vspace{-1em}
		\centerline{\includegraphics[scale=0.85]{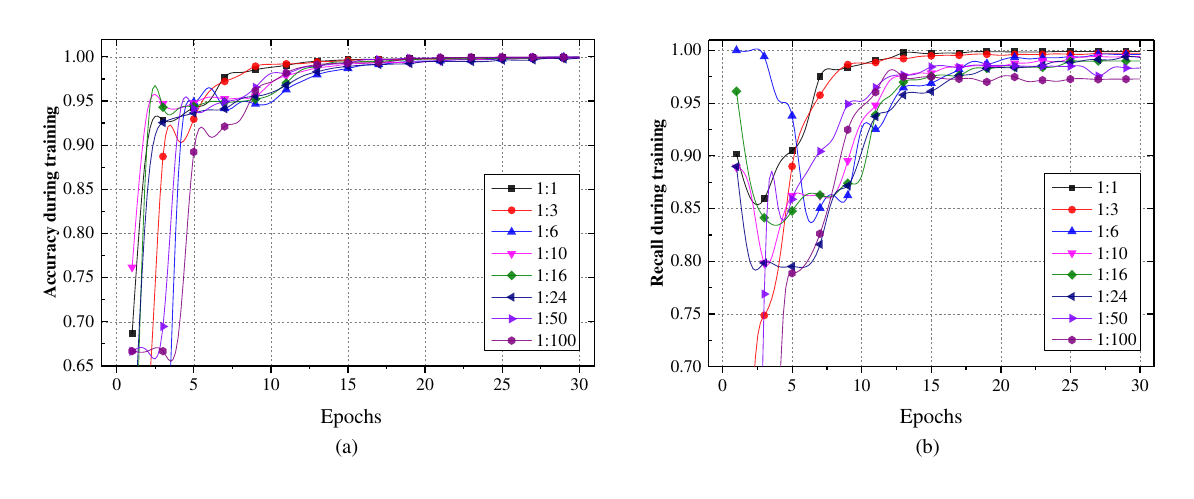}}
		\vspace{-1.5em}
		\caption{Training process of the model in different scenes (malicious : benign).}
		\label{roubustess}
		\vspace{-0.5em}
	\end{figure}
	
	In actual networks, the proportion of malicious traffic and benign traffic is extremely unbalanced, and most of the traffic is benign. Therefore, whether the model has good detection performance in a data imbalanced experimental environment is one of the keys to its application in real network environments. We design a variety of malicious : benign training scenarios with different proportions to test the robustness of the model. 
	
	The training process of the model is shown in Fig.~\ref{roubustess}. where figure (a) represents the change in accuracy of the model during convergence, and figure (b) represents the change in the recall of the model in the validation set during convergence. As the number of malicious samples in the training dataset decreases, the convergence rate of the model becomes slower, and the recall rate for malicious samples decreases. However, HSTF-Model has achieved convergence in these experimental scenarios, maintaining a detection accuracy of more than 97\%.
	
	\begin{table}[htbp]
		\vspace{-0.5em}
		\caption{Test results of HSTF-Model in different scenarios.}
		\vspace{-0.2em}
		\begin{center}
			\begin{tabular}{|c|c|c|c|}
				\hline
				\multirow{2}{*}{\textbf{malicious : benign}}&\multicolumn{3}{|c|}{\textbf{Evaluation index(\%)}}  \\
				\cline{2-4} 
				& \textbf{\textit{P}} & \textbf{\textit{R}} & \textbf{\textit{F1}} \\
				\hline
				\hline
				1 : 1 &  99.66&     99.28&     99.47  \\
				\hline
				1 : 3 &  99.76    & \textbf{99.74}&     99.75 \\
				\hline
				1 : 6 &  99.96 &     99.66    & \textbf{99.81}\\
				\hline
				1 : 10 &99.76 &     99.42&     99.59 \\
				\hline
				1 : 16 & 99.84 &     99.00    & 99.42\\
				\hline
				1 : 24 &  99.78 &     99.34&     99.56\\
				\hline 
				1 : 50 &  99.90 &     98.34&     99.11 \\
				\hline
				1 : 100 &  \textbf{99.98} &     97.30 &     98.62 \\
				\hline
			\end{tabular}
			\label{roubustess-tab}
		\end{center}
		\vspace{-1.5em}
	\end{table}
	
	The test results of the model in the dataset are shown in Tab~\ref{roubustess-tab}. The analysis shows that with the reduction of malicious samples in the dataset, the neural network is trained to fit more benign data and extract the characteristics of benign traffic. This makes the detection threshold of the model for malicious traffic increase, and the model is more and more inclined to judge the network behavior as benign. This means that the model is more biased towards the precision.	Only when the malicious characteristics of a flow are very obvious, the model will judge it as malicious.	When the number of malicious samples is reduced to a certain extent, the model will lose its ability to discriminate for malicious traffic. However, HSTF-Model has excellent robustness,  from 1:1 to 1:100, the F1 of model is 98.62\%$ \sim $99.81\% and the FPR 0.34\%$ \sim $0.02\%. Even when the ratio is 1:100, that is, when the malicious sample only accounts for about 0.99\% in training, the model can still achieve the precision rate of 99.98\%, the recall rate of 97.3\% and the F1 of 98.62\%.  
	
	\subsection{\textbf{Generalization analysis of HSTF-Model}}
	\subsubsection{\textbf{Generalization of HSTF-Model in different scenarios}}
	A method is considered to have good generalization performance when it can be used to detect other datasets with different distributions after learning in one dataset. In this paper, we train HSTF-Model in the dataset BTHT-2018 and use the dataset ISCX-2012 to verify its generalization.
	
	\begin{table}[htbp]
		\vspace{-1em}
		\caption{The generalization detection results of HSTF-Model in ISCX-2012 under different scenarios.}
		\vspace{-0.5em}
		\begin{center}
			\begin{tabular}{|c|c|c|c|c|c|}
				\hline
				\multirow{2}{*}{\textbf{malicious : benign} } &\multirow{2}{*}{\textbf{packet size}}&\multirow{2}{*}{\textbf{flow size}} &\multicolumn{3}{|c|}{\textbf{Evaluation index(\%)}} \\
				\cline{4-6} 
				& & &\textbf{\textit{P}} & \textbf{\textit{R}} & \textbf{\textit{F1}} \\
				\hline
				\hline
				1 : 1     &400 &    3 &    85.76 &    95.64      &90.43 \\
				\hline
				1 : 1     &400 &    6 &    77.19 &    96.02     &85.58 \\
				\hline
				1 : 1     &800 &    3 &    \textbf{91.41} &    95.72     &\textbf{93.51} \\
				\hline
				1 : 1 &    800&    6 &    80.69 &    96.44     &87.86 \\
				\hline
				1 : 10 &    400 &    3 &    83.52 &    94.98 &    88.88 \\
				\hline
				1 : 10 &    400     &6&    80.54 &    96.43 &    87.73 \\
				\hline
				1 : 10 &    800     &3    &87.03 &    95.72 &    91.17 \\
				\hline
				1 : 10 &    800     &6&    80.70 &    \textbf{96.52} &    87.91 \\
				
				\hline
			\end{tabular}
			\label{roc-tab}
		\end{center}
		\vspace{-1.5em}
	\end{table}
	
\begin{figure}[htbp]
		\vspace{-2em}
		\centerline{\includegraphics[scale=1.2]{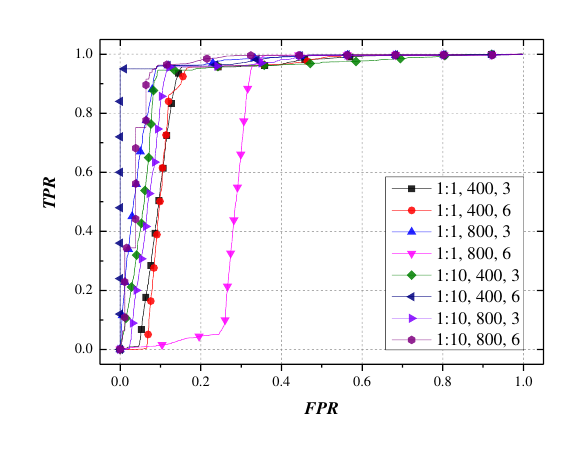}}
		\vspace{-2.5em}
		\caption{ROC curves of the model in ISCX-2012 under different scenarios. "1 : 1, 400, 3", for instance, means that the ratio of malicious : benign = 1 : 1 in training, packet size = 400 bytes and flow size = 3 packets.}
		\label{RoC}
		\vspace{-1em}
	\end{figure}

	The test results of the experiment are shown in Tab.~\ref{roc-tab}. When HSTF-Model is used to detect datasets with different distributions, the performance of model is degraded. When flow size = 3 and packet size = 800, the model can get the best generalization, that is, the precision = 91.41\%, the recall rate = 95.72\%, and the F1 = 93.51\% are obtained in ISCX-2012. At the same time, in order to better show the expected generalization performance of the model, we show the ROC curves of the model in different scenarios in Fig.~\ref{RoC}.
	
	\subsubsection{\textbf{Comparison with other methods}}
	A model usually pays more attention to generalization and hopes to obtain good detection performance in different experimental scenarios. Therefore, we compared HSTF-Model with other methods in generalization. In addition, there are no matched high-quality related papers in existing research work for HTTP-based Trojan malicious traffic detection during the on-line phase. Therefore, we select several machine learning methods that are widely used in the current intrusion detection field for comparison. Tian \emph{et al.}\cite{tian2017detection} used decision trees to conduct Android repackaged malware detection based on code heterogeneous analysis, and Senavirathne\emph{et al.}\cite{senavirathne2019integrally} used decision trees for privacy attack detection based on the intruder's uncertainty. Gu \emph{et al.}\cite{gu2019novel} proposed a support vector machine (SVM) integration framework for intrusion detection, which can get good robustness. Vijayanand \emph{et al.}\cite{vijayanand2018intrusion} proposed an SVM-based intrusion detection system based on a wireless mess network with high detection accuracy. Bost \emph{et al.}\cite{bost2015machine} used hyperplane decision (SVM, \textit{etc.}), Naive Bayes, and decision trees in the classification of encrypted data to construct a comprehensive classifier for detection.
	
	The experimental results are shown in Tab~\ref{other}. Obviously, HSTF-Model has the best comprehensive performance. Although Naive Bayes is 98.22\% in the recall rate, the accuracy rate is only 58.78\%. This means that the method will produce a large number of false positives during actual detection, which is unacceptable. In addition, the decision tree and SVM series methods trained in the dataset BTHT-2018 can obtain detection rate of 90+\% in BTHT-2018, but both show serious over-fitting phenomenon in dataset ISCX-2012, in which the SVM (linear) completely loses the detection ability in ISCX-2012. In general, other methods cannot detect the attack effectively.
	
	\begin{table}[htbp]
		\vspace{-0.5em}
		\caption{Generalization comparison of HSTF-Model with other methods in dataset ISCX-2012.}
		\vspace{-0.5em}
		\begin{center}
			\begin{tabular}{|c|c|c|c|}
				\hline
				\multirow{2}{*}{\textbf{Method} } &\multicolumn{3}{|c|}{\textbf{Evaluation index(\%)}} \\
				\cline{2-4} 
				&\textbf{\textit{P}} & \textbf{\textit{R}} &  \textbf{\textit{F1}} \\
				\hline
				\hline
				Naive Bayes     & 58.78 &    \textbf{98.22} &    73.50\\
				\hline
				Decision Tree (CART)      &    1.25 &    0.02 &    0.04\\
				\hline
				Decision Tree (C4.5)     &     11.36 &    0.10     &0.20\\
				\hline
				SVM (linear) &    0 &    0 &    0 \\
				\hline
				SVM (rbf) &0.11 &    0.06 &    0.10\\
				\hline
				HSTF-Model &\textbf{91.41} &    95.72     &\textbf{93.51}\\
				\hline
			\end{tabular}
			\label{other}
		\end{center}
		\vspace{-1em}
	\end{table}

	 In addition, we also compared the HSTF-Model with two non-hybrid neural networks (individual CNN and LSTM model) to verify the effectiveness of our hybrid structure. The experimental results are shown in Tab~\ref{other2}. Although CNN and LSTM are basically consistent with HSTF-Model in terms of recall, the precision is far lower than HSTF-Model. Therefore, the hybrid structure improves the ability of model to distinguish malicious traffic, which can effectively reduce false positives.
	
	\begin{table}[htbp]
		\vspace{-0.5em}
		\caption{Generalization comparison of HSTF-Model with individual CNN and LSTM in dataset ISCX-2012.}
		\vspace{-0.5em}
		\begin{center}
			\begin{tabular}{|c|c|c|c|}
				\hline
				\multirow{2}{*}{\textbf{Method} } &\multicolumn{3}{|c|}{\textbf{Evaluation index(\%)}} \\
				\cline{2-4} 
				&\textbf{\textit{P}} & \textbf{\textit{R}} &  \textbf{\textit{F1}} \\
				\hline
				\hline
				CNN & 78.14 & \textbf{95.72} & 86.04   \\
				\hline
				LSTM & 79.37 & 95.66 & 86.76   \\
				\hline
				HSTF-Model &\textbf{91.41} &   \textbf{ 95.72  }   &\textbf{93.51}\\
				\hline
			\end{tabular}
			\label{other2}
		\end{center}
		\vspace{-1em}
	\end{table}
	
	\section{Related work}
	The work of this paper is HTTP-based Trojan detection, which belongs to the field of intrusion detection. The method used is anomaly detection based on deep learning. In this section, we introduce related research work and detection methods.
	
	\subsection{\textbf{Intrusion detection systems}}
	Intrusion detection is an auxiliary method for network firewalls, which uses various technical methods to detect and alert network attacks. An intrusion detection system (IDS) is a network security product that uses intrusion detection methods to monitor the traffic flowing through the device, and issues an alert or takes proactive measures when the suspicious transmission is found. In an increasingly complex network environment, people often deploy IDSs to protect equipment and information. An effective intrusion detection technology is the core of IDSs. Trojans attack detection is an important supplement to IDSs.
	
	The methods used for intrusion detection can generally be divided into two categories, signature detection, and anomaly detection\cite{depren2005intelligent}. Signature detection, also called misuse detection, establishes a characteristic behavior model for known attacks. When a similar pattern is detected from the network, the corresponding behavior will be determined to be malicious. Signature detection can maintain a high accuracy rate for known attacks, but this method cannot detect a new type of unknown attack, that is, a 0-day attack, so its practical application has limitations and there are not many related studies. Anomaly detection, also called behavior detection, is one of the mainstream methods of intrusion detection\cite{bhuyan2013network}. It can detect 0-day attacks through feature analysis, which is very important for network security. Because the data in the network is constantly updated and new types of attacks are constantly generated, we need detection methods that can detect new types of attacks. In this paper, therefore, we focus on anomaly detection. Based on survey analysis, we divide it into classic anomaly detection, traditional machine learning-based (TML-based) anomaly detection, and deep learning-based (DL-based) anomaly detection.
	
	\subsection{\textbf{Classical anomaly detection}}
	Classical anomaly detection is the method used in early anomaly detection. For the benign network behavior, a corresponding behavior pattern feature library is established\cite{depren2005intelligent}. When the traffic is detected to deviate from the outline of benign network behavior, the system determines it to be abnormal\cite{can2015survey}.
	
	Classic anomaly detection methods can detect unknown attacks to a certain extent. However, it cannot achieve the same detection capability as feature detection when detecting known attacks\cite{kemmerer2002intrusion}. Furthermore, abandoning the modeling of malicious behaviors and learning only the current benign behaviors will make the model tend to judge all-new network behaviors as abnormal, which will lead to a high false-positive rate and make the model lose its discriminating ability\cite{depren2005intelligent}. Also, the completeness of feature engineering will greatly affect the detection performance of the model. Therefore, at present, researchers begin to pay more attention to combining signature detection and anomaly detection, while learning normal and abnormal network behaviors, and comprehensively extracting features to discriminate behaviors. Because this hybrid detection method mainly focuses on anomaly detection and can detect unknown attacks, it is also considered to be anomaly detection methods\cite{bhuyan2013network}.
	
	\subsection{\textbf{TML-based anomaly detection}}
	There are many anomaly detection methods that use traditional machine learning methods to detect malicious traffic\cite{buczak2015survey} anomaly detection. For instance, Bayes\cite{panda2007network}, Markov\cite{chen2016anomaly} and so on. In general, these anomaly detection methods learn both normal and abnormal characteristics, and can effectively detect both known attacks and new types of attacks.
	
	Mishra \emph{et al.}\cite{mishra2018detailed} conducted a detailed investigation and analysis of various machine learning technologies, compared the ability of machine learning technologies to detect attacks, looking for the cause of problems in machine learning when detecting intrusion behaviors. Aljawarneh \emph{et al.}\cite{aljawarneh2018anomaly} proposed a hybrid machine learning method for intrusion detection, which uses a voting algorithm to filter the data. The hybrid algorithm consists of J48, Meta Pagging, RandomTree, REPTree, AdaBoostM1, and so on. On binary and multi-class NSL-KDD datasets\cite{tavallaee2009detailed}, the accuracy of the model is 99.81\% and 98.56\%, respectively. Chen \emph{et al.}\cite{chen2018machine} proposed an unbalanced data gravity-based classification (IDGC) algorithm to classify unbalanced data for detecting malicious mobile applications. Gezer \emph{et al.}\cite{gezer2019flow} studied the use of machine learning to monitor banking Trojans. Using random forest classifiers in experiments can achieve 99.95\% accuracy. Al-Yaseen \emph{et al.}\cite{al2017multi} proposed an improved K-means algorithm for building high-quality training datasets, proposed a multilayer hybrid intrusion detection model and used support vector machines and extreme learning machines for detection. In the KDD-Cup’99 dataset\cite{rosset2000kdd}, 95.75\% accuracy can be achieved. Wang \emph{et al.}\cite{wang2020botmark} propose BotMark for botnets detection based on flow-based and graph-based network traffic behaviors. BotMark uses k-means to measure the similarity and stability of flows, uses the least-square technique and Local Outlier Factor (LOF) to calculate anomaly scores that measure the differences of their neighborhoods. Experimental results show that BotMark's detection accuracy reaches 99.94\%.
	
	There are also some shortcomings for TML-based anomaly detection. Most anomaly detection requires expert knowledge to design feature engineering. Feature engineering is first designed, and then supervised or unsupervised algorithms are used to build detection models based on these features. However, designing a high performance feature engineering that reflects the essential characteristics of data is a problem that is still an ongoing research issue, namely representation learning\cite{zhang2013effective}. Simultaneously, this detection method relies on feature engineering. The uncertainty of human factors and the limitations of the scene will prevent the detection model from achieving good robustness and generalization.
	
	\subsection{\textbf{DL-based anomaly detection}}
	Deep learning has also attracted widespread attention in the field of intrusion detection\cite{Yang2018Machine}. Many researchers have proposed intrusion detection methods based on deep learning. After a simple preprocessing of the network behavior, the neural network can automatically extract features from the data for updating parameters and can perform incremental learning. DL-based anomaly detection methods do not require researchers to pay more resources to establish feature engineering, which reduces the difficulty of experimental research. Because the model can automatically abstract features, the model can maintain good generalization as long as the data resources are satisfied\cite{dong2016comparison}. This is an advantage over other anomaly detection methods.
	
	Kwon \emph{et al.}\cite{kwon2017survey} summarized deep learning methods, introduced the latest research on deep learning technology with network anomaly detection as the core, and proved the feasibility of deep learning methods in network traffic analysis. Javaid \emph{et al.}\cite{javaid2016deep} studied the use of deep learning techniques to help system administrators detect network security vulnerabilities in organizations. In the high-dimensional problem domains of anomaly detection, Erfani \emph{et al.}\cite{erfani2016high} proposed a hybrid model that trains unsupervised DBN to extract general underlying features, which can effectively improve the detection speed. Zhou \emph{et al.}\cite{zhou2017anomaly} extended a deep autoencoder to eliminate outliers and noise. The superior performance of this method is proved on a series of benchmark problems. Shone \emph{et al.}\cite{shone2018deep} proposed an asymmetric deep autoencoder (NDAE) for unsupervised feature learning, which was evaluated on KDD-Cup'99 and NSL-KDD datasets. Li \emph{et al.}\cite{li2017intrusion} proposed an image conversion method for NSL-KDD data, using convolutional neural networks to automatically learn the features of the graph NSL-KDD transform. The results show that CNN is sensitive to image transformation of attack data and can be used for intrusion detection.
	
	In general, based on existing research, the following points can be improved in the field of Trojan attack detection: 1) The benchmark dataset (dataset KDD-Cup’99, \textit{etc.}) is becoming obsolete. As the network environment becomes more complex, new datasets are needed. 2) At present, there are is no specific work based on HTTP-based Trojan detection in the on-line phase. 3) Anomaly detection methods based on deep learning have advantages over other detection methods. However, most of the current detection methods directly use the raw traffic for learning and do not effectively use the artificial feature processing experience. In this paper, we propose a method for malicious traffic detection based on deep learning. In addition to the raw data, we also use expert knowledge to extract statistical features of spatio-temporal sequences for model training and detection, which makes the model have excellent detection performance in the actual network environment.
	
	\section{Discussion}
	\subsection{\textbf{Limitations of HSTF-Model}}
	HSTF-Model can learn the characteristics of traffic data well and can perform incremental learning. New data can be input to the model iteratively, and the decision function of the model is continuously updated, so that the model can resist concept drift to a certain extent.
	
	However, the model still has some shortcomings. First, if the flow data of a sample is small and the corresponding extractable features are small, the model is prone to misjudgment, but this is reasonable. Secondly, although the generalization of HSTF-Model is good enough compared with other methods, there is still a 5\% -6\% decrease in detection performance on different datasets, which is a problem that we need to improve in the future. In addition, we have not fully verified the performance of the model on encrypted traffic. But HSTF-Model extracts information from the statistical characteristics of the traffic. Therefore, we believe that the model can also get good results on encrypted traffic in the basis of sufficient training.
	
	\subsection{\textbf{Limitations of experimental design and datasets}}
	In this paper, we determine some optimal parameters through experiments. Because there are many combinations of parameters, we cannot implement and evaluate all the schemes, but only select some local optimums. As a rule of thumb, we use control variables and distributed values to cover the optimal solution as much as possible.
	
	For the number of  malicious traffic.  In the actual network environment, the proportion of HTTP-based Trojan malicious traffic is very low. We captured about 37,000 HTTP-based Trojan flows from the actual network environment. We admit that this number occupies a small proportion in the BTHT-2018 dataset, but we believe that this amount of data can guarantee to meet the experimental needs. Experiments show that the HTTP-based Trojan data we captured can roughly cover the behavioral characteristics of such attacks. 
	
	For the content of experimental data, there are few or even single packets in many flows, and the behavior cycle time is short. This is because: 1) The attacker's server cannot receive the request (the domain name expires and is disabled, the server is shut down, \textit{etc.}). 2) The response may be blocked by other detection systems. 3) The flow itself is a one-way message transfer. However, experiments show that HSTF-Model still has excellent detection ability for such data.
	
	\section{Conclusion}
	In this paper, we propose a spatio-temporal sequence feature model to describe HTTP-based Trojan attacks, and build a prototype detection method HSTF-Model to detect HTTP-based Trojan traffic based on deep learning. Also, we collected the traffic from the actual network to generate the dataset BTHT-2018. Experiments show that the combination of raw data and statistical features can more fully display the inherent characteristics of the traffic, and neural networks can more fully learn the data. HSTF-Model fed with the dataset BTHT-2018 can reach F1 of 99.47\% in the same dataset, and it can also reach 93.51\% in the public dataset ISCX-2012 (20+\% improvement compared with other methods), which proves that the generalization performance of the model is excellent, while other traditional methods do not have generalization. In addition, HSTF-Model can obtain the F1 value of 98.62\% in the 1 : 100 dataset, proving that the model has excellent robustness. 
	
	In the future, we will take the improvement and expansion of the BTHT-2018 dataset as one of the main tasks, and provide more complete data scenarios. Simultaneously, it is planed that the fine-grained multiple classification will be performed in order to detect more types of network attacks.

	
	
	
	\bibliographystyle{elsarticle-num}

	\bibliography{sample}
	
\end{document}